\documentclass[twocolumn,showpacs,preprintnumbers,showkeys]{revtex4}

\usepackage{amsmath}
\usepackage{amsfonts}
\usepackage{amssymb}
\usepackage{graphicx}
\allowdisplaybreaks[4]
\usepackage{natbib}
\begin{document}


\newcommand{\abs}[1]{\ensuremath{\left\lvert #1\right\rvert}}
\newcommand{\erf}[2][]{\ensuremath{\text{erf#1}\left(#2\right)}}
\newcommand{\kB}{\ensuremath{k\!_B}}
\newcommand{\eps}{\ensuremath{\varepsilon}}
\newcommand{\lD}{\ensuremath{\lambda_{\text D}}}

\newcommand{\be}{\begin{equation}}
\newcommand{\ee}{\end{equation}}
\newcommand{\bs}{\begin{subequations}}
\newcommand{\es}{\end{subequations}}
\def\ee{\end{equation}}
\def\be{\begin{equation}}
\def\bdm{\begin{displaymath}}
\def\edm{\end{displaymath}}
\def\mgf{\vec{B}}
\def\eef{\vec{E}}
\def\ep{\epsilon }
\def\vv{\vec{v}}
\def\xx{\vec{x}}
\def\kk{\vec{k}}
\def\kks{\vec{k^{'}}}
\def\kpa{k_{\parallel }}
\def\kper{k_{\perp }}
\def\jnn{J_n(z)}
\def\jns{J_n^{\prime }(z)}
\def\jn2{J_n^2(z)}
\def\udf{{\cal U}f_a^{(0)}}
\def\wdf{{\cal W}f_a^{(0)}}
\def\vdf{\vec{V}f_a^{(0)}}
\def\vpa{v_{\parallel }}
\def\vper{v_{\perp }}
\def\Omm{\Omega }
\def\ppa{p_{\parallel }}
\def\pper{p_{\perp }}
\def\ppv{\vec{p}}
\def\px{p_{\perp }}
\def\pz{p_{\parallel }}
\def\omm{\omega }
\def\noi{\noindent}
\def\l{\left}
\def\r{\right}
\def\a{\alpha }
\def\om{\omega }
\def\rp{\omega _{nR}}
\def\omr{\omega _R}
\def\Oma{\Omega _a}
\def\yu{Y_{L,U}}
\def\Om{\Omega _p}
\def\wa{\omega _{p,a}}
\def\wpe{\omega _{p,e}}
\def\wpa{\omega ^2_{p,a}}
\def\Tp{\Theta _p}
\def\Te{\Theta _e}
\def\Tn{\Theta }
\def\upa{u_{a,\parallel}}
\def\ut{u_{a}}
\def\2kua{\sqrt{2}ku_{a,\parallel}}
\def\Zst{Z^{'}}
\def\bp{\beta _{\parallel }}
\def\w{w^2}
\def\wi{{\mu +1\over \mu }w^2}
\def\a{\alpha }
\def\d{\delta }
\def\wp{weakly propagating }
\def\ct{cyctronic }
\def\mpm{\mu ^{1/2}}
\def\mi2{\mu ^{-1/2}}
\def\m32{\mu ^{3/2}}
\def\ea{\epsilon _a}
\def\an{\alpha _0}
\def\bc{\beta _c}
\def\os{\omm _{\sigma }}
\def\La{\Lambda }
\def\la{\lambda }
\def\pw{\partial _{\omm }}

\newcommand{\R}{\ensuremath{\mathds{R}}}

\newcommand{\pa}{\ensuremath{_\shortparallel}}
\newcommand{\se}{\ensuremath{_\perp}}

\newcommand{\Th}{\ensuremath{\varTheta}}
\newcommand{\Ga}{\ensuremath{\varGamma}}

\newcommand{\uint}{\ensuremath{\int_{-\infty}^\infty}}

\newcommand{\f}[1]{\ensuremath{\boldsymbol{#1}}}
\newcommand{\im}{\ensuremath{\mathfrak{I\!m}}}

\newcommand{\pd}[2][]{\ensuremath{\frac{\partial #1}{\partial #2}}}
\newcommand{\df}{\ensuremath{\mathrm{d}}}

\newcommand{\etal}{ \emph{et\,al.}}

\title{Cosmic magnetization: from spontaneously emitted aperiodic turbulent to ordered equipartition fields}
\author{R. Schlickeiser$^{1,2}$}
\email{rsch@tp4.rub.de}
\affiliation{1 Institut f\"ur Theoretische Physik, Lehrstuhl IV:
Weltraum- und Astrophysik, Ruhr-Universit\"at Bochum, D-44780 Bochum, Germany\\
2 Research Department Plasmas with Complex Interactions, Ruhr-Universit\"at Bochum, D-44780 Bochum}
\date{\today}

\begin{abstract}
It is shown that an unmagnetized nonrelativistic thermal electron-proton plasma spontaneously emits 
aperiodic turbulent magnetic field fluctuations of strength $|\delta B|=9\beta _eg^{1/3}W_e^{1/2}$ G, 
where $\beta _e$ is the normalized thermal electron temperature, $W_e$ the thermal plasma energy density and $g$ the plasma parameter. Aperiodic modes fluctuate only in space, but are not propagating. For the unmagnetized intergalactic medium, immediately after the reionization onset, the field strength from this mechanism is about $4.7\cdot 10^{-16}$ G, too weak to affect the dynamics of the plasma. The shear and/or compression of the intergalactic medium exerted by the first supernova explosions amplify these seed fields and make them anisotropic, until the magnetic restoring forces affect the gas dynamics at ordered plasma betas near unity.
\end{abstract}


\maketitle

The interstellar medium (ISM) is filled with (1) a dilute mixture of charged particles, atoms, molecules and dust grains, referred to as
interstellar gas and dust, (2) partially turbulent magnetic fields, (3) dilute photon radiation fields from stars, dust and the universal microwave background radiation, and (4) cosmic ray particles with relativistic energies. It is known for a long time\cite{bd60,p66,p68}, even before the discovery of the universal cosmic microwave background radiation, that these ISM components have comparable energy densities and pressures, each of the order of $10^{-12}$ erg cm$^{-3}$, commonly referred to as the equipartition condition in the ISM. Since today, this truly remarkable equipartition in the ISM has not been understood nor explained theoretically. One refers to pressure partition, if the ratio of any two individual pressures is a constant, and to pressure equipartition, if this ratio is near unity. 
 
In other astrophysical objects equipartition conditions, and the closely related minimum-energy assumption, for the total magnetic field energy density and the kinetic energy density of plasma particles are also often invoked for convenience\cite{l92} in order to analyze cosmic synchrotron intensities. The minimum-energy assumption was first proposed by Burbidge\cite{b56} and applied to the optical synchrotron emission of the jet in M87. Duric\cite{du90} argued that any major deviation from equipartition would be in conflict with radio observations of spiral galaxies. Observationally, for a variety of nonthermal sources 
the equipartition concept is supported by magnetic field estimates as e.g. in the Coma cluster of galaxies 
and radio-quiet active galactic nuclei\cite{s87}. Also the solar wind plasma exhibits near equipartition conditions: 
ten years of Wind/SWE satellite data\cite{mar11} near 1 AU showed that the proton and electron temperature anisotropies $A=T_\perp/T_\parallel $ are bounded by the mirror and firehose instabilities at large values of the parallel plasma beta $\beta _\parallel =8\pi nk_BT_\parallel /B^2\ge 1$ and by parallel propagating Alfven waves\cite{smi11} at small values of $\beta _\parallel <1$, resulting in near magnetic field equipartition. 

Because of their comparably low gas densities, all cosmic fully and partially ionized non-stellar plasmas are collision-poor, as indicated by the very small values of the plasma parameter $g=\nu _{ee}/\wpe \le 10^{-10}$, given by the ratio of the electron-electron Coulomb collision frequency $\nu _{ee}$ to the electron plasma frequency $\wpe $, characterizing interactions with electromagnetic fields, so that fully kinetic plasma descriptions are necessary. Because of the large sizes of astrophysical systems compared to the plasma Debye length, the fluctuations are descibed by real wave vectors ($\vec{k}$) and complex frequencies $\omega (\vec{k})=r(\vec{k})+\imath \gamma (\vec{k})$, 
implying for the space- and time-dependence of e.g. magnetic fluctuations the superposition of 
$\delta \vec{B}(\vec{x},t)\propto \exp [\imath (\vec{k}\cdot \vec{x}-rt)+\gamma t]$. 
One distinguishes between collective modes with a fixed frequency-wavenumber dispersion relation, also referred to as normal modes, and non-collective (no frequency-wavenumber relation) modes in the system. Regarding frequency, basically two fundamental types of fluctuations occur: (1) weakly amplified/damped solutions (e.g. Alfven waves, electromagnetic waves) with $\vert \gamma \vert \ll \omega _R$, and (2) weakly propagating solutions (e.g. firehose and mirror fluctuations) with $\omega _R\ll \gamma $, including aperiodic solutions with $\omega _R=0$ (e.g. Weibel fluctuations\cite{w59}). Aperiodic modes fluctuate only in space, do not propagate as $r=0$, but permanently grow or decrease in time depending on the sign of $\gamma $. Past research\cite{sa60,s67,is73,k98} has concentrated predominantly on the fluctuations from collective weakly amplified modes in the plasma.
 
All plasmas, including unmagnetized and those in thermal equilibrium, have fluctuations so that their state variables such as density, pressure and electromagnetic fields fluctuate in position and time. Unlike for weakly amplified/damped modes, however, for aperiodic fluctuations the expected fluctuation level has never been calculated quantitatively. Only recently general expressions for the electromagnetic fluctuation spectra (electric and magnetic field, charge and current densities) from uncorrelated plasma particles in unmagnetized plasmas for arbitrary frequencies have been derived\cite{sy12} using the system of the Klimontovich and Maxwell equations, which are appropriate for fluctuations wavelengths longer than the mean distance between plasma particles, i.e. $k\le k_{\rm max}=2\pi n_e^{1/3}$. These general expressions are covariantly correct within the theory of special relativity, and hold for arbitrary momentum dependences of the plasma particle distribution functions and for collective and non-collective fluctuations. The electric\cite{fo1} and magnetic field fluctuations in unmagnetized plasmas with the plasma frequency $\wpa =(4\pi e^2n_a/m_a)^{1/2}$ 

\be
\begin{pmatrix} <\delta E^2_\parallel  >_{k,\om} \\
<\delta E^2_\perp >_{k,\om} \\ 
<\delta B^2>_{k,\om} \\
\end{pmatrix}=
\sum_a{\wpa m_a\over 4\pi ^3k^2}
\begin{pmatrix} {K_{\parallel  }(k, \om )\over
                |\om \Lambda _L(\kk ,\om )|^2} \\ 
                {K_{\perp }(k, \om )\over
                |\om \Lambda _T(\kk ,\om )|^2} \\
                {c^2k^2K_{\perp }(k, \om )\over
                |\om ^2\Lambda _T(\kk ,\om )|^2} \\
\end{pmatrix}
\label{c1}
\ee
are given in terms of the parallel and perpendicular form factors  

\be
\begin{pmatrix} K_{\parallel }(k, \om ) \\
K_{\perp }(k, \om )\end{pmatrix}
=k^2\Re \int d^3p{f_a(\ppv )\over \gamma
+\imath (\kk \cdot \vv -r)}
\begin{pmatrix} \vpa ^2 \\ \vper ^2 \end{pmatrix}
\label{c2}
\ee
and the general longitudinal and transverse dispersion functions $\Lambda _{L,T}(\kk ,\om)$ ivolving the respective parallel and perpendicular dielectric tensor elements. Eqs. (\ref{c1}) -- (\ref{c2}) are the generalizations of the standard expression found in the literature in which the weak.amplification limit of $\gamma\to 0^+$ is taken at the outset to approximate the factor $-\imath (\gamma +\imath (\kk \cdot \vv -r))^{-1}$ by $\lim_{\gamma\to 0^+}(-\imath)(\gamma +\imath(\kk \cdot \vv -r))^{-1}\to\pi\;\delta(\kk \cdot \vv -r)$. 

We now consider the unmagnetized intergalactic medium (IGM) immediately after the reionization onset, assuming that any earlier cosmological magnetization has vanished during the long recombination era with a fully neutral IGM. Modeling the photoionization by 
the first forming stars\cite{hg97,hh03} indicates IGM temperatures of about $T_e=T_p=T=10^4T_4$K and ionized gas densities of 
$n_e=10^{-7}n_7$ cm$^{-3}$ at redshift $z=4$. For this isotropic thermal IGM proton-electron plasma we follow recent work\cite{sy12} to  calculate from Eqs. (\ref{c1})- Eqs. (\ref{c2}) the spontaneously emitted magnetic field fluctuation spectrum of aperiodic ($r=0$)
fluctuations 

\bdm
(4\pi ^{5/2})<\delta B^2>_{k,\gamma }=
\edm
\be
\sum_a{\wpa m_a\ut c^2kD\bigl({\gamma \over k\ut }\bigr)\over 
[\gamma ^2+c^2k^2+\pi ^{1/2}\sum_a{\wpa |\gamma |\over k\ut }
D({\gamma \over k\ut})|]^2},
\label{c3}
\ee
with the thermal velocity $\ut =\sqrt{2k_BT_a/m_a}$ and $D(x)=e^{x^2}\hbox{erfc }(|x|)$ denoting  
the complimentary error function. The related collective Weibel mode\cite{w59} has a positive growth rate in anisotropic plasma distribution functions, but is not excited in isotropic plasma distributions.  

Integrating over all values of $\gamma $ and $k$ provides the energy density of spontaneously emitted fully random magnetic fluctuations 

\be
<\delta B^2>=4\pi \int_0^{k_{\rm max}} dk\, k^2\, <\delta B^2>_{k}
\label{c4}
\ee
with 

\bdm
<\delta B^2>_{k}={\wpe ^2m_ec^4k^2\beta _e^2\over 4\pi ^{5/2}}
\edm
\be
\times \int_{-\infty }^\infty dx\, {F(x,\mu )\over 
\bigl[k^2c^2(1+\beta _e^2x^2)+\pi ^{1/2}\wpe ^2|x|F(x,\mu )\bigr]^2},
\label{c5}
\ee
where 

\be
F(x,\mu )=D(x)+\mu ^{-1}D(x\mu ),
\label{c6}
\ee
the mass ratio $\mu ^2=m_p/m_e=1836$ and $\beta _e=u_e/c=1.84\cdot 10^{-3}T_4^{1/2}$. 
Likewise, one finds for aperiodic fluctuations no charge density and parallel electric field fluctuations 
$<\delta \rho ^2>_{k}=<\delta E^2_\parallel  >_{k}=0$, and 

\bdm
<\delta E^2_\perp >_{k}={\wpe ^2m_ec^4k^2\beta _e^4\over 4\pi ^{5/2}}
\edm
\be
\times \int_{-\infty }^\infty dx\, {x^2F(x,\mu )\over 
\bigl[k^2c^2(1+\beta _e^2x^2)+\pi ^{1/2}\wpe ^2|x|F(x,\mu )\bigr]^2}
\label{c7}
\ee
The symmetric function $D(-x)=D(x)$ attains its maximum value $D(0)=1$ at $x=0$ and is monotonically decreasing to $0$ as $x\to \infty$. Its rational approximation better than $2.5\cdot 10^{-5}$ is given by\cite{as72} $ D(x)\simeq a_1t-a_2t+a_3t^3$ %
with $t=/(1+px)$, $p=0.47047$, $a_1=0.3480242$, $a_2=0.0958798$ and $a_3=0.7478556$. 
Given the smallness of $a_2$ we use as lower and upper limits $D_L(x)< D(x)<D_U(x)$ with $D_L(x)=(a_1-a_2)t+a_3t^3$ and $D_U(x)=a_1t+(a_3-a_2)t^3$, i.e. 

\be
D_L(x)\simeq (a_0+a_3t^2)t,\;\;\; D_U(x)\simeq (a_1+a_4t^2)t
\label{c8}
\ee
with 
$a_0=a_1-a_2=0.2521444$ and $a_4=a_3-a_2=0.6519758$. Fig. 1 indicates that the relative deviation of the upper and lower limits are smaller than 30 percent at all values of $x$. 

\begin{figure}
\begin{center}
\includegraphics[width=84mm]{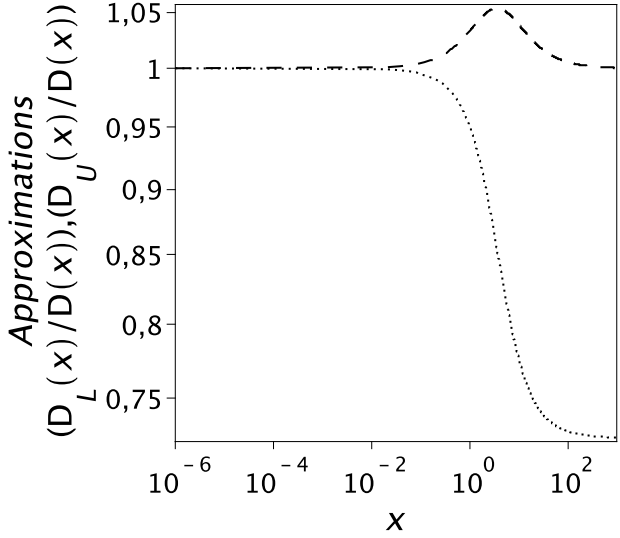}
\end{center}
\caption{Relative variations of the lower ($D_L(x)/D(x)$) (dotted curve) and upper ($D_U(x)/D(x)$) (dashed curve) limits  as a function of $x$.} 
\end{figure}
Because of the large value of $\mu =43$ we can neglect the proton contribution in Eq. (\ref{c5}) so that $F(x,\mu )\simeq D(x)$, implying in terms of the normalized wave vector $\kappa =kc/\wpe $ that 

\be
<\delta B^2>_{k}={m_ec^2\beta _e^2\over 2\pi ^{5/2}\kappa ^2}J(\beta _e,\kappa )
\label{c9}
\ee
where 

\be
J(\beta ,\kappa )=\int_0^\infty dx\, {D(x)\over \bigl[1+\beta ^2x^2+{\pi ^{1/2}\over \kappa ^2}xD(x)\bigr]^2}
\label{c10}
\ee
Eq. (\ref{c4}) then becomes 

\be
<\delta B^2>={2\over \pi ^{3/2}}m_ec^2\beta _e^2\bigl({\wpe \over c}\bigr)^3
\int_0^{{2\pi cn_e^{1/3}\over \wpe}}d\kappa \, J(\beta _e,\kappa )
\label{c11}
\ee
The integral (\ref{c10}) is well approximated by $J_L(\beta ,\kappa )<J(\beta ,\kappa )<J_U(\beta ,\kappa )$ 
with 

\be
J_{U,L}(\beta ,\kappa )=\int_0^\infty dx\, {D_{U,L}(x)\over \bigl[1+\beta ^2x^2+{\pi ^{1/2}\over \kappa ^2}xD_{L,U}(x)\bigr]^2}
\label{c12}
\ee
After straigtforward but tedious algebra we derive  

\bdm
{J_{L,U}(\beta ,\kappa )\over a_{0,1}/(2p)}\simeq {1\over (1+\yu )^2}\Bigl({\yu ^2-2\yu -1\over \yu (1+\yu )}
\edm
\be
+{3\yu ^2+2\yu +1\over \yu ^2}\ln \bigl(1+\yu \bigr)+2\ln {p\over \beta }\Bigr)
\label{c13}
\ee
with $\yu (\kappa )=\pi ^{1/2}a_{1,0}/[p\kappa ^2]$. The asymptotic expansions for small and large values of $\yu $ correspond to large and small values of the normalized wavenumber, respectively, providing 

\bdm
J_{L,U}(\beta ,\kappa )\simeq {a_{0,1}\over 2p}
\edm
\be
\times 
\begin{cases}
2\ln (p/\beta )+{1\over 2} & \text{for $ \yu \ll 1 $ } \\
{1+3\ln \yu +2\ln (p/\beta )\over \yu ^2} & \text{for $ \yu \gg 1 $ } 
\end{cases}
\label{c14}
\ee
The resulting expressions for $<\delta B^2>_k^{L,U}$ from Eq. (\ref{c5}) to leading order increase $\propto \kappa ^4$ at small normalized wavelength $\kappa \le (a_{1,0}\pi ^{1/2}/p)^{1/2}$ and approach constants at large $\kappa $, disagreeing with earlier results\cite{y07,ts07} which were based on fluctuation spectrum formula valid for weakly damped/amplified modes. The constants at large values of $\kappa $ provide the dominating contribution to the remaining $\kappa $-integral in Eq. (\ref{c11}). We find $<\delta B^2>_{U}=1.38<\delta B^2>_{L}$ with 

\bdm
<\delta B^2>_{L}={4a_{0}\over \pi ^{1/2}p}\ln \bigl({pe^{1/2}\over \beta _e}\bigr)m_ec^2\beta _e^2n_e^{1/3}\bigl({\wpe \over c}\bigr)^2
\edm
\bdm
=5{\beta _e^2\over \bigl[{4\pi \over 3}n_e\lambda ^3\bigr]^{2/3}}8\pi n_ek_BT=
5\beta _e^2g^{2/3}8\pi n_ek_BT
\edm
\be
={80\pi \over 3}\beta _e^2g^{2/3}W_e=2.2\cdot 10^{-31}T_4n_7^{4/3}\; \hbox{erg cm}^{-3}, 
\label{c15}
\ee
with the thermal energy density $W_e=3n_ek_BT/2=2.1\cdot 10^{-19}n_7T_4$ erg cm$^{-3}$ and the plasma beta $g=2.3\cdot 10^{-13}n_7^{1/2}T_4^{-3/2}$. This magnetic field energy density corresponds to a minimum total fluctuating magnetic field strength of 

\be
|\delta B|_L=\bigl(<\delta B^2>_{L}\bigr)^{1/2}=4.7\cdot 10^{-16}T_4^{1/2}n_7^{2/3}\; \hbox{G}
\label{c16}
\ee
This fluctuating magnetic field strength is still small enough to allow rectilineal propagation of ultrahigh energy cosmic ray protons in the IGM, 
The associated turbulent plasma beta is larger than 

\be
\beta _t={8\pi n_ek_BT\over <\delta B^2>_{U}}={1\over 6.9\beta _e^2g^{2/3}}=
1.1\cdot 10^{13}n_7^{-1/3}
\label{c17}
\ee
Now we finally address how ordered magnetic field structures emerge from these randomly distributed magnetic fields. As we demonstrated the unmagnetized, isotropic, thermal and steady IGM plasma by spontaneous emission generates steady tangled fields, isotropically distributed in direction, on small spatial scales $\le 10^{10}n_7^{-1/2}$ cm (corresponding to $\kappa \ge 1$).   
Because of its ultrahigh turbulent plasma beta value (\ref{c17}), these seed fields are too weak to affect the dynamics of the IGM plasma, but are tied passively to the highly conducting IGM plasma as frozen-in magnetic fluxes. Earlier analytical considerations and numerical simulations\cite{hoe79,la80,haa85,ms90,la02} showed that any shear and/or compression of the IGM medium enormously amplify these seed magnetic fields and make them anisotropic. Considering a cube containing an initially isotropic magnetic field, which is compressed to a factor $\eta \ll 1$ times its original length along one axis ($z$), these authors showed that the perpendicular magnetic field components are enhanced by the factor $\eta ^{-1}$. Depending on the specific exerted compression and/or shear, even one-dimensional ordered magnetic field structures can be generated out of the original isotropically tangled field configuration\cite{la02}. Hydrodynamical compression or shearing of the IGM medium arises from the shock waves of the supernova explosions of the first stars at the end of their lifetime, or from supersonic stellar and galactic winds.  The IGM seed magnetic field upstream of these shocks is random in direction, and by solving the hydrodynamical shock structure equations for oblique and conical shocks it has been demonstrated\cite{cc90,c06}, that the shock compression enhances the downstream magnetic field component parallel to the shock, but leaving the magnetic field component normal to the shock unaltered. Consequently, a more ordered downstream magnetic field structure results from the randomly oriented upstream field. Such stretching and ordering of initially turbulent magnetic fields is also seen in the numerical hydrodynamical simulations of supersonic jets in radio galaxies and quasars\cite{ms90}. 

This passive hydrodynamical amplification and stretching of magnetic fields continues until the magnetic restoring forces affect the gas dynamics, i.e. at ordered plasma betas near unity. During these stretchings and amplifications also MHD dynamo processes need to be considered\cite{bs05}.  The stretching and ordering of magnetic fields also affects the motions of the IGM protons and electrons, so that our original expressions for the spontaneously emitted fluctuation spectra (\ref{c1}) in unmagnetized plasmas then no longer apply. 

In principle, our suggested mechanism of spontaneously emitted aperiodic turbulent magnetic fields should also operate during earlier cosmological epochs before recombination. However, the then relativistic temperatures of the unmagnetized plasma require the use of the relativistic fluctuation spectra which also are available\cite{sy12}. 

I gratefully acknowledge helpful discussions with Drs. P. H. Yoon, A. Beresnyak and M. Lazar.


\begin{thebibliography}{99}
\bibitem{fo1}We corrected a typo in the expresssion for $<\delta E_{\parallel }^2>_{k,\omm}$.
\end{thebibliography}
\end{document}